\documentclass[10pt,conference]{IEEEtran}

\ifCLASSOPTIONcompsoc
  \usepackage[nocompress]{cite}
\else
  \usepackage{cite}
\fi

\usepackage{amssymb}
\usepackage{booktabs}
\usepackage{cite}
\usepackage{enumitem}
\usepackage[pdftex]{graphicx}
\usepackage{lscape}
\usepackage[framemethod=tikz]{mdframed}
\usepackage{rotating}
\usepackage{tabularx}
\usepackage{tikz}
\usepackage{tipa}
\usepackage{xtab}
\usepackage[hidelinks]{hyperref}
\usepackage{bookmark}

\usepackage[]{todonotes}

\begin{document}

\setlist{nolistsep}

\title{ADEPT: A Socio-Technical Theory\\ of Continuous Integration}
\author{
	\IEEEauthorblockN{Omar Elazhary}
	\IEEEauthorblockA{
		University of Victoria\\
		omazhary@uvic.ca
	}
	\and
	\IEEEauthorblockN{Margaret-Anne Storey}
	\IEEEauthorblockA{
		University of Victoria\\
		mstorey@uvic.ca
	}
	\and
	\IEEEauthorblockN{Neil A. Ernst}
	\IEEEauthorblockA{
		University of Victoria\\
		nernst@uvic.ca
	}
	\and
	\IEEEauthorblockN{Elise Paradis}
	\IEEEauthorblockA{
		University of Toronto\\
		Elise.Paradis@utoronto.ca
	}
}

\maketitle
\IEEEpeerreviewmaketitle

\begin{abstract}
Continuous practices that rely on automation in the software development workflow have been widely adopted by industry for over a decade. 
Despite this widespread use, software development remains a primarily human-driven activity that is highly creative and collaborative.
There has been extensive research on how continuous practices rely on automation and its impact on software quality and development velocity, but relatively little has been done to understand how automation impacts developer behavior and collaboration.
In this paper, we introduce a socio-technical theory about continuous practices. The ADEPT theory combines constructs that include humans, processes, documentation, automation and the project environment, and describes propositions that relate these constructs.
The theory was derived from phenomena observed in previous empirical studies.
We show how the ADEPT theory can explain and describe existing continuous practices in software development, and how it can be used to generate new propositions for future studies to understand continuous practices and their impact on the social and technical aspects of software development.
\end{abstract}



%

\section{Introduction}
\label{sec:intro}

Continuous practices in software development are motivated by the premise that committing small but frequent changes to software repositories will minimize integration problems and provide rapid feedback to developers \cite{fowler2006continuous, humble2010continuous}.
This premise has extended over the years from continuous integration (CI) to include both the release (continuous delivery - CDE) and deployment (continuous deployment - CD) processes, among others \cite{fitzgerald2017continuous, fowler2006continuous, humble2010continuous}.

Automation has been shown to have a positive effect on the development process in terms of merge cadence and software quality \cite{vasilescu2015quality}.
However, while there is an extensive body of research that focuses on automation, its optimization, and its effects on the software development process, there has been little done to investigate its impact on software developers.
This neglect is concerning because software development is first and foremost a human-driven activity.
Software requirements are generated by clients (humans), formulated into features by product managers (humans), designed and implemented by developers (humans), tested by developers/testers (humans), and released to the aforementioned clients.
To be clear, standards, practices, and tools play an essential role in each of these processes, but with the goal of supporting human effort.
That is, they cannot substitute for the creative and human-centric aspects as human decision making is still what drives the development process forward.

While research on software development automation and human interaction is sparse, there is more understanding of how humans interact with automation in other domains.
For example, the work of Parasuraman et al. considers how automation affects decision making in a generic context. \cite{parasuraman1993performance, parasuraman1997humans, parasuraman2010complacency}.
They highlight three main phenomena that occur when humans interact with automation; abuse, misuse, and disuse.
\emph{Abuse} refers to the overuse of automation such that it becomes more of hindrance than an asset.
\emph{Misuse} refers to the over-reliance on automation in such a way that the human rarely questions results the automation produces.
\emph{Disuse} refers to the opposite of misuse and occurs when a human distrusts all results automation produces.
These, and possibly other generic automation phenomena, influence the relationship between humans and automation and likely transfer to software developers when interacting with automated build and test tools.

In our effort to study the relationship between human, automation, and process within the context of CI, we \textbf{built an explanatory theory that represents the relationships between humans, automation, and the software development process} that is based on previous empirical studies that we and others have conducted.
Our proposed theory, ADEPT (\textbf{A}utomation, \textbf{D}ocumentation, \textbf{E}nvironment, \textbf{P}rocess, \textbf{T}eam member), captures and explains several relationships that exist between developers, automation, and CI.
It can also be used to reason about existing phenomena and explore new human-focused propositions that have gone unnoticed by studies of system data alone (as this data does not always capture the human aspects that our theory presents).
The ADEPT theory is a step towards a more theory-oriented software engineering research field, as advocated by Stol and Fitzgerald \cite{stol2015theory}.
We hope it will spark human- and theory-oriented discussions, investigations, and ideas with respect to continuous practices.

\section{Missing Humans in the CI Research Context}
\label{sec:background}

CI practices have been widely adopted in the past decade partly because of their flexibility when it comes to adapting to evolving requirements in today's fast-paced software development market \cite{fitzgerald2017continuous}.
CI facilitates rapid developer feedback from both their testing tools and from customers because changes are integrated, tested, and deployed faster and more frequently \cite{fowler2006continuous, beck1999embracing}.

To enable rapid feedback, automation is typically introduced as a way of speeding up the process \cite{fowler2006continuous}.
A CI server (such as TravisCI\footnote{\url{https://travis-ci.org}} or Jenkins\footnote{\url{https://www.jenkins.io}}) is now a common addition to development environments.
Such tools are configured by developers to run a variety of tasks which can be repeatable.
For instance, they can be configured to build an application, test it and report on test results, or package and deploy it to a target system \cite{staahl2017cinders}.

The topic of CI and its purported benefits is a well-studied topic in the software research literature.
However, the majority of studies focus on either the impact of CI on the development process, or the nature of the automation applied in the CI process.
With respect to the process, Vasilescu et al. \cite{vasilescu2015quality} found that open-source projects using TravisCI experience an increase in change merge frequency and a decrease in number of bugs opened.
Similar results were later reported by both Hilton et al. \cite{hilton2016usage} and Zhao et al. \cite{zhao2017impact}.

The literature has also considered anti-patterns in CI, which are incorrect ways of using a feature of automation.
Duvall \cite{duvall2011continuous} defined a dictionary of CI process anti-patterns, which Zampetti et al. augmented \cite{zampetti2020empirical}.
Literature focused on automation anti-patterns has attempted to automatically discover and suggest fixes for anti-patterns in tool configuration, such as the work done by Vassallo et al. \cite{vassallo2019automated} and Gallaba et al. \cite{gallaba2018use}.

There is a large body of research that focuses on automation and process but does not consider the human developer in any capacity beyond configuring recommender systems and smell detectors.
We are only aware of three studies that investigate human-related phenomena within the context of CI.
Gupta et al. \cite{gupta2017impact} investigated the impact that adopting a CI tool has on the ability of an open source project to attract and retain developers.
They found that both attractiveness and retention decreased once TravisCI was adopted by a project, however, it is unclear why.
Pinto et al. \cite{pinto2017inadequate} studied the role a CI tool plays in a developer's impression of their development process.
While it increases developer confidence in their code state, they find that sometimes there is a false sense of confidence when developers blindly run tests, which echoes the phenomenon of \emph{misuse} mentioned in section \ref{sec:intro}.
Souza and Silva \cite{souza2017sentiment} used sentiment analysis of commit messages to investigate the relationship between developer mood and the likelihood of broken builds.
They found that negative commit messages are more likely to cause broken builds, which in turn result in negative commits attempting to fix them.

While not as frequently investigated as the relationship between automation and process, the previous three studies indicate that automation has an impact on the developers using it.
This impact is an under-explored relationship between developers and automation that could severely affect the development process. 
Automation and process are only successful to the extent human developers adopt or buy in to the tool or process. 
The nature of the relationship between developers and CI tools and process is unclear.
\section{The ADEPT Theory}
\label{sec:theory}
Over the past few years, we have been studying how software developers perceive CI practices and automation, and how they interact with automation in a CI context \cite{elazhary2019not,elazhary2020tse}.
The constructs and propositions we present in our theory emerged from two empirical studies.
Our first study investigated how process documentation was described in open source contribution files, and whether automation and CI practices were part of this documentation \cite{elazhary2019not}.
The second investigated how CI practices and automation are integrated in software development workflows from a human perspective, and their impact on the entire software development process \cite{elazhary2020tse}.
The theory we introduce has also been informed by a review of other CI studies.

\begin{figure*}
	\centering
	\includegraphics[width=1.7\columnwidth]{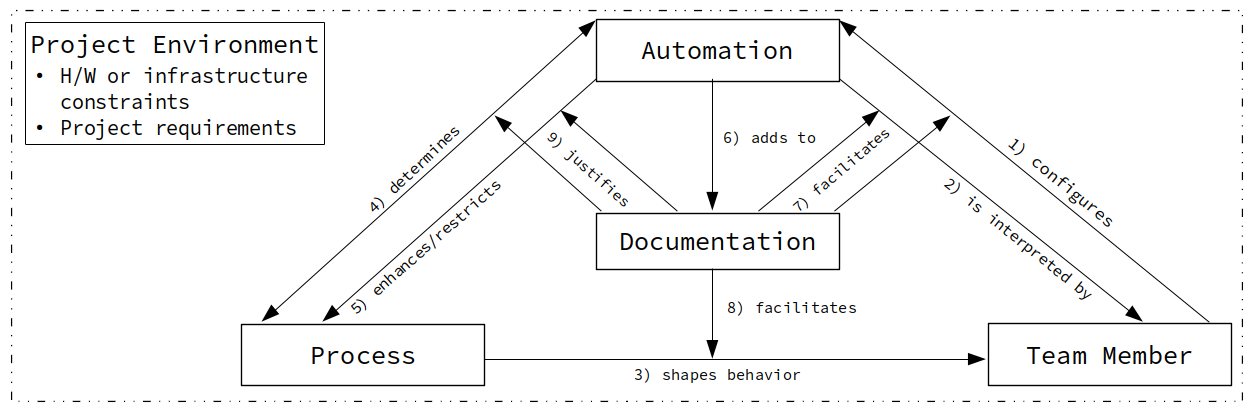}
	\vspace{-4mm}
	\caption{Visual representation of the ADEPT theory (boxes are constructs, edges are propositions)}
	\label{fig:theory}
	\vspace{-5mm}
\end{figure*}

The name of our theory, ADEPT, stands for Automation, Documentation, Environment, Process, Team member. 
At its core, this is a meso-level theory \cite{creswell2017research} that aims to explain the relationship between human behavior and automation without ignoring the role the development process plays.
It is explanatory in the sense that it highlights relationships between its different constructs, some of which have not been fully explored (as discussed in Section \ref{sec:background} \cite{varpio2019distinctions}).
A visual representation of the theory is given in Figure \ref{fig:theory}.

\subsection{Constructs}
The main constructs in our theory are as follows.

\subsubsection{Automation}
This construct refers to any type of automation that triggers when a developer/team member makes a change to a software project.
It is an abstraction of the automation described by Humble and Farley \cite{humble2010continuous} and an example of which (TravisCI) was studied by Beller et al. \cite{beller2017travistorrent}.
Examples of this include the jobs that run on a commit or pull/merge request.
It can also include the tools that run within these jobs, such as dependency management tools, build tools, testing tools, or bots that act in the background.

\subsubsection{Team Member}
This construct is the core of the ADEPT theory.
It captures the different types of humans that interact with and are affected by automation as well as the software development process.
Based on existing literature \cite{gousios2015work, gousios2016work, steinmacher2015systematic, humble2018accelerate}, we know that software developers and reviewers are important examples of team members we need to consider in studies of CI.
Our previous studies \cite{elazhary2019not,elazhary2020tse} identified additional team members that interact within or are affected by the CI process: team leads, DevOps engineers, and product managers.
Understandably, product managers will have limited influence on the tool and will often not be in a position to configure it, unlike DevOps engineers.
Team leads have an interesting relationship with both automation and process.
A team lead has a bird's-eye perspective of how the entire process works.
Depending on their duties, a team lead may also interact with automation as a reviewer, and occasionally, in a development capacity.

\subsubsection{Process}
This construct captures the series of steps/activities a feature goes through from its inception as a requirement/issue until it is released (delivery) or deployed to a production system (deployment).
This construct is intended as an abstract representation of the development process, though due to the prevalence of the pull request model \cite{gousios2014exploratory} in practice, we assume the process follows a similar structure/workflow.

\subsubsection{Documentation}
This construct captures externalized information available to developers that pertains to the project they are working on.
Most notably, it captures artifacts created solely for the purposes of documenting processes/tools (README or contribution guides \cite{prana2019categorizing, elazhary2019not}) as well as those that document development activity on features (feature/issue documentation, pull/merge request documentation, commit documentation) and how they are connected.
It may also capture tacit knowledge (typically kept in a developer's head) in the form of explicit quasi-documentation artifacts that give insight into how to perform project-related activities, such as build scripts (sometimes called infrastructure-as-code) \cite{bartusevics2015models}.

\subsection{Propositions}
The following propositions link the constructs described above and represent both phenomena from the two studies we conducted \cite{elazhary2019not,elazhary2020tse}, as well as phenomena observed in other studies.

\subsubsection{Team Member Configures Automation}
\label{rel:config}
A software developer or DevOps engineer (and rarely a product manager) is responsible for setting up and configuring the supporting project infrastructure.
They are directly responsible for the configuration complexity of the resulting scripts.
This proposition captures where we may observe instances of automation abuse (over-automation with no discernible positive effect) \cite{parasuraman1997humans} or the introduction of configuration smells \cite{gallaba2018use}.

\subsubsection{Automation Is Interpreted by Team Member}
\label{rel:interpret}
This proposition depicts the effect of the automation tool/mechanism on the software developer's decision making process when deciding to integrate code into the codebase.
Here we might see cases of misuse (placing too much trust in the automation) or disuse (not trusting the automation enough) \cite{parasuraman1997humans}.
An example of disuse is a reviewer ignoring build notifications because the automation notifications arrive too frequently on several channels \cite{elazhary2020tse}.

\subsubsection{Process Shapes Team Member Behavior}
\label{rel:shape}
A team member is expected to adhere to a particular process when performing their tasks.
This process is typically part of their team culture and should be documented somewhere.
Therefore, it can be said that the process (i.e., the agreed-upon way of doing things) guides a team member to behave a certain way \cite{elazhary2019not}.
As a prescribed example, in a project that adopts a CI process, developers are expected to commit in small increments, and commit often.
As a reactionary example, we might see cases of program managers using feature toggles as a throttling mechanism to cope with the flurry of seemingly unrelated, small changes.
Another example is reviewers being forced to perform manual usability testing \cite{elazhary2020tse} because the automation cannot perform this type of testing to the reviewers' satisfaction.

\subsubsection{Process Determines Automation}
\label{rel:determine}
The structure of the development process influences a team's choice of tools and the automation required.
This proposition may capture justification for the use of particular tools and reasons for automation configuration \cite{elazhary2019not}, and it is typically mediated by project context (hardware and infrastructure constraints, as well as the nature of project requirements).
However, it is also possible that tool choice may determine process, and that a workflow or process be constructed or adapted around the desire to use a particular tool, which may indicate a case of abuse in terms of how and why certain automation tools are adopted \cite{parasuraman1997humans}.

\subsubsection{Automation Enhances/Restricts Process}
\label{rel:enh/res}
The most common reason for introducing automation into a project's workflow is to enhance its speed, reduce cognitive overhead of executing manual steps, increase quality, or improve the onboarding process \cite{vasilescu2015quality, fowler2006continuous, fitzgerald2017continuous}.
It is possible, however, that automation may introduce negative effects, such as bottlenecks in build queues or long build times due to automation configuration complexity \cite{duvall2011continuous}.

\subsubsection{Automation Adds to Documentation}
\label{rel:adddoc}
Another phenomenon is the use of automation configuration and execution scripts as a way to externalize previously tacit information.
Developers and newcomers could consult such artifacts to familiarize themselves with the steps required to build a project \cite{elazhary2020tse} and use this to aid in their understanding of the process.

\subsubsection{Documentation Facilitates a Team Member's Interpretation of an Automation}
\label{rel:docinterpret}
Depending on how automation jobs are configured and the type of output they produce, it is possible that documentation is required to interpret the result of a job.
We consider two types of documentation: tool documentation that facilitates configuration and provides general tool orientation to new users, and infrastructure-as-code documentation which we discussed previously.

\subsubsection{Documentation Shapes a Team Member's Behavior in Following a Process}
\label{rel:docshape}
README files and contribution guides in open source projects typically attempt to document how (and in some cases why) a developer needs to perform various tasks in order to contribute to a project \cite{prana2019categorizing, elazhary2019not}.
This proposition is also identified as a solution to barriers facing newcomers in open source projects \cite{steinmacher2015systematic}.

\subsubsection{Documentation Justifies How Automation Enhances/Restricts a Process}
\label{rel:docenh/res}
Documentation (or infrastructure-as-code) gives us a glimpse into what choices are made to either enhance or restrict a process.
Choices made by the member who configured the automation tool/mechanism should give valuable insight into why the process functions the way it does and the perceived effects associated with automation.

\section{ADEPT Theory Application Examples}
\label{sec:instance}

To demonstrate the ADEPT theory's explanatory power and utility, we use the constructs and propositions we described above  to explain existing empirical observations.
For instance, the study conducted by Vasilescu et al. \cite{vasilescu2015quality} finds that the presence of automation in open source projects is correlated with fewer bugs and a higher merge rate.
Both of these results are represented by proposition 5, \emph{Automation Enhances/Restricts Process}, specialized in this case to `Enhance'.
However, what is not apparent (but highlighted by ADEPT) is what impact the adoption of TravisCI has on the developers, both directly in terms of decision making, and indirectly in terms of process changes that result in behavioral changes, with
the latter case being more important because productivity in terms of merging code is a function of human behavior as opposed to tool adoption.
Another example is the solution proposed by St{\aa}hl et al. \cite{staahl2017achieving} that focuses on establishing traceability between the software development process's artifacts (including automation), thereby strengthening documentation.
Stronger documentation justifies the success software developers report in their application of CI practices, as represented by propositions 5 and 9.
However, it is not clear how better documentation has affected developer behavior, nor how they perceive the automation.
Is it still a tool that they use to streamline their workflow, or has it evolved to become a record-keeping mechanism?

Considering human aspects, the work done by Pinto et al. \cite{pinto2017inadequate} posits a false sense of confidence that developers experience when blindly trusting tests.
This phenomenon represents misuse, which is illustrated by proposition 2.
However, it is unclear whether this misuse occurs because of the automation itself, or because of how it was configured (proposition 1), or even because of the restrictions the process imposes on it (proposition 4).
Another example is that of Souza and Silva's work on developer sentiment and how it is affected by broken builds \cite{souza2017sentiment}.
This phenomenon is represented by proposition 2 in that a broken build has a negative influence on the developer's next commit, which is more likely to cause another broken build.
That is, is the broken build the only factor contributing to worse contributions, or is build configuration another factor if it produces useless results, which makes fixing the problem an exercise in trial and error?
Or could it be that the process enforces a ``do not break the build" mentality that---when coupled with broken builds---leads to frustration, and thereby negative commit messages?
\section{Theory Validity}
\label{sec:validity}

In the previous section, we discussed two of six theory evaluation criteria proposed by Sj{\o}berg et al. \cite{sjoberg2008building}: utility and explanatory power.
Below we discuss the remaining four. 

\textbf{Testability}, also known as falsifiability, represents the extent to which a theory can be refuted using empirical evidence.
Our theory constructs represent the different entities that interact with each other within a software development ecosystem.
Each construct emerges and can be traced to previous empirical studies.
However, the propositions we discuss are based on our observations within a specific context.
For instance, we encountered a case where switching to a different automation (in that case Kubernetes) involved changes to the process itself (automation determines process) \cite{elazhary2020tse}.
These process changes entailed changes in developer behavior (process shapes team member behavior) and resulted in a challenging transition period as the automation impeded their move to CI.
While we observed this phenomenon in one organization, other organizations may experience a smoother transition experience.
In this case, context has a direct impact on falsifiability.

\textbf{Empirical power} refers to the theory's ability to explain why a phenomenon occurs.
This is typically fulfilled by two criteria: analogy and explanatory breadth.
To comply with the analogy criterion, we have made sure (where possible) to use existing empirical studies as the basis for our constructs and propositions.
With respect to explanatory breadth, we have (above) used the theory to position existing empirical studies and explain why their observed phenomena occurred.

\textbf{Parsimony} is a measure of theory minimalism, in the sense that it does not contain any unnecessary constructs or propositions.
Since this theory focuses on human interaction with automation within the context of continuous practices, we have elected to avoid introducing software code constructs.
We made this choice for two reasons.
The first is that considerations of the project code introduce an additional dimension of complexity, with all of its affiliated constructs and propositions.
The second is because project code can vary vastly from one project to the next, which would have forced us to saddle the theory with an extensive set of assumptions, thereby robbing the theory of its necessary generality.

\textbf{Generality} reflects how wide of a scope a theory has and to what extent it is independent of setting.
The theory we propose is an emerging work that is still in its infancy, but the constructs and propositions emerged from our earlier studies (conducted in industry settings) and a literature review.
Furthermore, we demonstrated its applicability to four phenomena that occur in both open source projects and industry in the previous section.
We do not claim the theory is transferable in its entirety across contexts, but we hope it is general enough to be valuable to studies of CI in other contexts.
\section{Concluding Remarks}
\label{sec:conclusion}

Previous studies have predominantly investigated the impact CI has on development process efficiency, or as part of a framework to establish how it impacts overall processes in an organization \cite{staahl2017achieving, fitzgerald2017continuous}.
Fewer studies consider its impact on human aspects (exceptions include \cite{gupta2017impact} and \cite{pinto2017inadequate}, as mentioned in Section~\ref{sec:intro}).
The lack of attention on human aspects is unfortunate as software development is a human-driven activity that is augmented by automation and continuous practices, not replaced by them. 
The ADEPT theory helps situate findings of existing studies and reveal gaps, while the theory propositions further suggest new studies on how CI impacts team members, process, and automation, and the pivotal role documentation plays.

\ifCLASSOPTIONcompsoc
   The Computer Society usually uses the plural form
  \section*{Acknowledgments}
\else
  \section*{Acknowledgment}
\fi

We acknowledge the support of the Natural Sciences and Engineering Research Council of Canada (NSERC).
We would also like to thank Cassandra Petrachenko for her valuable help with this study.

\ifCLASSOPTIONcaptionsoff
  \newpage
\fi

\bibliographystyle{IEEEtran}
\bibliography{main}

\end{document}